# ON THE GENERALIZED BOHM SHEATH CRITERION IN DUSTY PLASMAS

K.ŻUCHOWSKI (WARSZAWA)

*Institute of Fundamental Technological Research, PAS, Świętokrzyska 21, 00-049, Warsaw, Poland.*

The paper analyses Bohm criterion for the formation positive ion sheath in dusty plasmas . The criterion may be derived from the dispersion relation.

## 1. Introduction

Bohm considered the criterion for the formation of a stable positive ion sheath originally [1] and later then modified to a more general form [2]. Allen [3] was shown that the criterion could be derived from the dispersion relation. The sheath criterion has been applied to many applications such as probes and ion sources. However, this holds for the case where electrons follow the Maxwellian velocity distribution. The plasma often deviates from a thermally equilibrium state. In such a case, it is necessary to obtain the criterion and the sheath edge potential from the measured energy distribution. Amemiya [4] received the criterion for the formation of a stable positive ion sheath for an arbitrary electron velocity distribution function to obtain the ion flux. Next, he applied this criterion to plasmas containing negative ion [5].

Plasmas with dust grains are of interest both for the cosmic space as well as for the laboratory plasmas. Examples include cometary environments, planetary rings, the instellar medium, and the earth's magnetosphere [6]. Dust has been found to be a determinant component of rf plasmas used in the microelectronic processing industry, and it may also be present in the limiter regions of fusion plasmas due to the sputtering of carbon by energetic particles. It is interesting to note the recent flurry of activity in dusty plasma research. It has been driven largely by discoveries of the role of dust in quite different settings: The ring of Saturn [6] and the plasma processing device [7]. Dusty plasmas contain, beside positive ions and electrons, big particles usually negatively charged. They are conglomerations of the ions, electrons and neutral particles. These big particles, to be called grains, have atomic numbers $Z_d$ in the range of $10^4$-$10^6$ and their mass $m_d$ can be equal to $10^6$ of protons mass or more. It is assumed, for simplicity, that all grains have equal masses and charges, which are practically constant, and the collisions of all particles are neglected in the considered interval of time. We note that for dusty plasmas ratio of electrical charges of grains to it masses is usually much smaller than in the case of negative ions [8]. The ratio of charge to mass for a given component of the plasma determines its dynamics [9].

In considered dusty plasmas the size of grains is small to compare with average distance between grains. Because dynamics of the dusty plasma (electrons, positive ions and dusty grains) are quite different in the time and length scales here considered, then the equations for aforementioned components of the dusty plasma may be different. We use fluid and Vlasov descriptions for appropriate components of the dusty plasma.

## 2. Generalized sheath criterion

It is well known that for the Maxwellian velocity distribution for electrons the ion velocity $u_i$ required for the stable sheath formation is given by the ion acoustic speed [1] as

$$(1) \qquad C_s = \left(k_B T_e / m_i\right)^{1/2},$$

where: $k_B, T_e$ and $m_i$ denote Boltzmann`s constant, temperature of electrons and mass of ions, respectively. This so-called Bohm criterion was later rewritten more rigorously as [2]:

$$(2) \qquad <u_i^{-2}> \equiv \int_{-\infty}^{\infty} \frac{f_i(u)}{u^2} du = \frac{m_i}{k_B T_e},$$

where $f_i(u)$ is the ion velocity distribution and $< >$ means the ensemble average of the argument. The relation can also be derived from the dispersion relations in which Allen showed that the dispersion of a longitudinal wave propagating with $\omega/k \to 0$ ($\omega$ is the frequency and $k$ is the wave number ) to agree with the sheath criterion eq. (2) in the limit of $k\lambda_{De} \to 0$, where $\lambda_{De}$ is the Debye length for the electrons. Equations (1) and (2) are not mathematically correct as they where obtained by series expansions. The simplifying by using the dispersion relation is not rigorous as it is based on the linearized Vlasov equations. The use of the dispersion relation is a test for interpreting the sheath criterion. It is a physical interesting to treat the sheath as a kind of a static wave and furthermore convenience in including the effect of velocity distribution. Due to this reason in [4] was finding the sheath criterion for plasmas with negative ions. We try to find the sheath criterion for plasmas with a dust along this line.

For a general case including electrons, positive ions and dust with the velocity distributions $f_e(u)$, $f_i(u)$ and $f_d(u)$ respectively, we have the dispersion relation in front of the sheath edge as

$$(3) \quad \omega_{pe}^2 \int_{-\infty}^{\infty} \frac{f_e(u)}{(\omega-ku)^2} du + \omega_{pi}^2 \int_{-\infty}^{\infty} \frac{f_i(u)}{(\omega-ku)^2} du + \omega_{pd}^2 \int_{-\infty}^{\infty} \frac{f_d(u)}{(\omega-ku)^2} du = 1,$$

where $\omega_{pe}$, $\omega_{pi}$ and $\omega_{pd}$ are plasma frequencies of electron, positive ion and dust respectively. The integral is performed along the Landau's contour [5] avoiding the singularity at $u = \omega/k$. The velocity distributions $f_e(u)$, $f_i(u)$ and $f_d(u)$ are normalized such that the integration from $-\infty$ to $\infty$ is unity. Due to considering the steady state, we put $\omega = o$. Rearranging the both side of eq. (3) by using $<v_e^2>$ (the mean square velocity of electrons), we obtain

$$(4) \frac{(1-Z_d\alpha)}{m_e} \int_{-\infty}^{\infty} \frac{f_e(u)du}{u^2} + \frac{1}{m_i} \int_{-\infty}^{\infty} \frac{f_i(u)du}{u^2} + \frac{Z_d\alpha}{m_d} \int_{-\infty}^{\infty} \frac{f_d(u)du}{u^2} = \frac{(k\lambda_D)^2}{[m_e <v_e^2>/2]},$$

where $m_e, m_i$ and $m_d$ are masses of electron, positive ion and grain of dust respectively, $\alpha$ is the ratio of the dust density $n_{d0}$ to the positive ion density $n_{i0}$ at the sheath edge. The equality $3m_e <v_e^2>/2$ corresponds to the mean energy $E_e:(3/2)k_BT_e$ in the case of Maxwellian velocity of the electron distribution. Although $f_e(u)$ is not always Maxwellian, we define the Debye length formally as $\lambda_D \equiv \omega_p^{-1} <v_e^2/2>^{1/2} = \omega_p^{-1}(E_e/3m_e)^{1/2}$, where $\omega_p^2 = n_{i0}e^2/\varepsilon_0 m_e$, and $\varepsilon_0$: dielectric constant in vacuum. This Debye length is the shielding length for an electron by positive ions. For obtaining the Debye length $\lambda_{Di}$ for positive ions in Maxwellian plasma,

Poisson's equation should be solved around a positive ion whose charge is screened by electrons and negative charged grains of dust:

$$\frac{1}{r^2}\partial\left(r^2\frac{\partial\phi}{\partial r}\right)/\partial r = \frac{-n_{i0}e}{\varepsilon_0}\left[1-(1-Z_d\alpha)\exp\left(\frac{-e\phi}{k_BT_e}\right)-Z_d\alpha\exp\left(\frac{-e\phi}{k_BT_d}\right)\right].$$

Expanding the equation for small $\phi$ and assuming $\phi = \exp(-r/\lambda_{Di})/r$, we obtain $\lambda_{Di} = \lambda_D[(1-Z_D\alpha)+\gamma Z_D\alpha]^{-1/2}$, where $\gamma$ being the ratio of the mean energy of electrons to that of grains of dust. In any cases of $\alpha$ and $\gamma$, it becomes $\lambda_{Di} \leq \lambda_D$.

The charge neutrality holds at the sheath edge as

(5) $$n_{e0} + Z_d n_{d0} = n_{i0}.$$

Due to this reason and by neglecting the imaginary part in (3) and (4), we obtain

(6). $$\frac{1}{m_i}<v_i^{-1}> = -\frac{(1-Z_d\alpha)}{m_e}\int_{-\infty}^{\infty}\frac{f_e(u)du}{u^2} - \frac{Z_d^2\alpha}{m_d}\int_{-\infty}^{\infty}\frac{f_d(u)du}{u^2} + \frac{(k\lambda_D)^2}{[m_e<v_e^2>/2]}.$$

It will be assumed that $k$ corresponds to the reciprocal of the characteristic length which is given by the sheath thickness and much longer than $\lambda_D$. Under this condition, we may put $k\lambda_D \to 0$ in eq. (6). Due to $\lambda_{Di} \leq \lambda_D$, $k\lambda_{Di} \to 0$ is automatically filled.

If we assume that $f_e(v)$ can be measured by plane probes, in the form energy distributed $F(E)$, where $E = m_e v^2/2$, as shown in [4], [5], than the first and the second term of right hand side of eq. (6) in the limit $k\lambda_D \to 0$ leads:

(7) $$\frac{1}{m_i}<v_i^{-2}> = (1-Z_d\alpha)\int_0^\infty\frac{F(E)}{2E}dE + Z_d\alpha\int_0^\infty\frac{F_d(E)}{2E}dE.$$

In eq. (7) a similar expression is used for a grain of dust velocity distribution function $f_d(v)$ and energy distribution $F_d(E)$.

Therefore, due to the acceleration by a weak electric field in front of the sheath, positive ions at the sheath edge have a velocity distribution with a narrow velocity width. In such a case [10],[11],[12] :

$$\frac{\omega}{k} = \left[\frac{n_{i0}}{n_{e0}}\left(\frac{k_B T_e}{m_i}\right)\right]^{1/2}.$$

Where: $\omega, k, n_{i0}$ and $n_{e0}$ denote a frequency of the wave, the wave number, unperturbed ion's density and unperturbed electron's density, respectively. These dust-plasma waves admit the case when the temperature of ions $T_i$ and the temperature of electrons $T_e$ are nearly equal: $T_i \approx T_e$. The presence of dust grains can have profound influence on low frequency waves leading to the phase velocity of the waves much smaller than thermal velocity of electrons and ions but much greater than the thermal velocity of the dust. For that dust-acoustic waves (DAW), in a longitudinal approximation, we have the following phase velocity [11]:

$$\omega / k = \beta v_d,$$

where:
$\beta^2 = Z_d(\delta - 1)/(1 + \gamma\delta)$, $v_d^2 = k_B T_e / m_d$, $\delta = n_{i0}/n_{e0}$ and $\gamma = T_e/T_i$.

For the above described dust-plasma waves ( DAW ) [11] and dust-plasma-ion (DIAW) [10], fluid approach was used. The same waves were obtained from the Vlasov-Poisson system of equation [12]. In both cases the global charge neutrality was assumed: $n_{i0} = Z_d n_{d0} + n_{e0}$, where $n_{d0}$ denote the unperturbed dust density.

### References.

1. D. BOHM, E.H. BURHOP and H.S. MASSEY, *The Characteristic of Electrical Discharges in Magnetic Fields*, A. Guthrie and R.K. Wakerling [Eds.], McGraw-Hill, New York 1949.
2. E. R. HARRISON and W.B. THOMPSON, *Low pressure plane symmetric discharge,* Proc. Phys. Soc., **74,** 145 ,1959.

3. J. E. ALLEN, *A note on the generalized sheath criterion,* J. Phys., D **9,** 2331, 1976.
4. H. AMEMIYA, *Sheath Formation Criterion and Ion Flux for Non-Maxwellian Plasma,* J. Phys. Soc. Jpn., **66,** 1335**,** 1997**.**
5. H. AMEMIYA., *Sheath Formation Criterion and Ion Flux for a Non-Maxwellian Plasma Containing Negative Ions,* J. Phys. Soc. Jpn., **67,** 1998.
6. G. K. GEORTZ, *Dusty plasmas in the solar system,* Rev. Geophys. **27**, 271, 1989.
7. R. L. MERLINO, A. BARKAN, C. THOMPSON and N. D`ANGELO, *Laboratory studies of waves and instabilities in dusty plasmas*, Phys. Plasmas, **5,** 1607, 1998**.**
8. S. BABOOLAL, R. BHARUTHRAM and M. A. HELMBERG, *Arbitrary-amplitude theory of ion-acoustic solitons in warm multifluid plasmas,* J Plasma, Phys., **41**, 341, 1989.
9. N. A. KRALL. and A. W. TRIVELPIECE, Principles of plasma physics, McGraw Hill, New York 1973
10. P. K. SHUKLA and V. P. SILIN, *Dust Ion-Acoustic Wave*, Physica Scripta, **45,** 508, 1992.
11. P. K. SHUKLA, *Low-frequency Modes in Dusty Plasmas,* Physica Scripta, **45**, 504, 1992.
12. A. J. TURSKI, B. ATAMANIUK and K. ŻUCHOWSKI, *Dusty plasma solitons in Vlasow plasmas,* Arch. Mech., **51**, 167, 1999.